\documentclass[amsmath, aps]{revtex4}
\usepackage{graphicx}
\usepackage{times}
\usepackage{epsfig}
\usepackage{amssymb}
\usepackage{amsmath}

\newcommand{\hr}{\hat{\bf r}}

\begin{document}
\title{Extension of the sum rule for the transition rates between multiplets
to the multiphoton case}

\author{D. Solovyev$^1$, L. Labzowsky$^{1,2}$, A. Volotka$^{1,3}$, and G. Plunien$^3$}

\affiliation{ 
$^1$ V. A. Fock Institute of Physics, St. Petersburg
State University, Petrodvorets, Oulianovskaya 1, 198504,
St. Petersburg, Russia
\\
$^2$  Petersburg Nuclear Physics Institute, 188300, Gatchina, St.
Petersburg, Russia
\\
$^3$ Institut f\"ur Theoretische Physik, Technische Universit\"at Dresden,
Mommsenstrasse 13, D-01062, Dresden, Germany}

\begin{abstract}
The sum rule for the transition rates between the components of two multiplets,
known for the one-photon transitions, is extended to the multiphoton transitions
in hydrogen and hydrogen-like ions. As an example the transitions $3p-2p$, $4p-3p$
and $4d-3d$ are considered. The numerical results are compared with previous
calculations.
\end{abstract}
\maketitle

\section{Introduction}
Many-photon processes were considered many times by different authors for
various atomic systems. The spontaneous two-photon decay in hydrogen atoms and
hydrogen-like ions was studied since the theoretical formalism has been introduced
by G\"oppert-Mayer \cite{Goppert} and the first evaluation for the two-photon
E1E1 transition $2s\rightarrow 2\gamma(E1)+1s$ has been presented by Breit
and Teller \cite{Breit}. A highly accurate calculation of the E1E1 transition
probability has been performed by Klarsfeld \cite{Klarsfeld}.

A larger number of transitions between $ns$ and $nd$ states was evaluated
in \cite{Tung}, \cite{TungTang}. Evaluation of E1E2 and E1M1 transition rates
for the $2p\rightarrow \gamma(E1)+\gamma(E2)+1s$ and
$2p\rightarrow \gamma(E1)+\gamma(M1)+1s$ processes have been first accomplished
in \cite{LabShon}, \cite{LabShonSol} for hydrogen-like ions within the wide
range of the nuclear charge $Z$ values: $1\leqslant Z\leqslant 100$.
Such calculations were performed within the fully relativistic approach.
For the summation over the intermediate states (i.e. over the complete Dirac
electron spectrum) the relativistic B-spline approach
\cite{sapirstein:1996:5213}
was used. In \cite{LabSol} these transition rates were evaluated in the
nonrelativistic limit employing the analytic expression for the nonrelativistic
Coulomb Green function \cite{RZM}. Later in \cite{Amaro} the relativistic
calculations for these transition rates were repeated and extended to some
other transitions. The results of  all the works \cite{LabShon}, \cite{LabShonSol},
\cite{LabSol}, \cite{Amaro} are in the reasonable agreement with each other.
The results in \cite{LabSol} can be compared with the others only for
small $Z$ values. In \cite{TransRev} the influence of the external electric
field on the two-photon transition rates in hydrogen and anti-hydrogen atoms
was studied; in the same work the three-photon $2p\rightarrow 3\gamma(E1)+1s$
transition rate was evaluated.

Recently the theory of the two-photon transition in hydrogen became very important
for the astrophysics. The interest was stimulated by the new accurate measurements
of the different properties of the cosmic microwave background \cite{Hin}, \cite{Page}.
These properties depend on the cosmological history of the hydrogen recombination.
The bound-bound one-photon transitions did not permit the atoms to reach their ground
state: each photon released by one atom was immediately absorbed and reemitted by
a neighbouring atom. As it was first established in \cite{Zeld}, \cite{Peebles},
the two-photon $2s-1s$ transition represents one of the main channels for the radiation
to escape the interaction with the matter. From this escaped radiation the cosmic
microwave background was later formed.

In \cite{Dubrovich}, \cite{Wong} it was argued that the $ns\rightarrow 1s$ ($n>2$)
and $nd\rightarrow 1s$ two-photon transitions can also give a sizable contribution
to the process of decoupling of the radiation from the interaction with the matter.
Recently this problem was investigated thoroughly in the theoretical astrophysical
studies in \cite{J.Chluba}, \cite{Hirata}. There is a crucial difference between
the decay of the $ns$ ($n>2$) or $nd$ levels and the $2s$ decay level. This
difference is due to the presence of the cascade transitions as the dominant decay
channels in case of $ns$ ($n>2$) and $nd$ levels. For the $2s$ level the cascade
transitions are absent. Since the cascade photons can be effectively reabsorbed,
the problem of separation of the ``pure'' two-photon contribution from the cascade
contribution arises. An interference between the two decay channels should also
be taken into account. This problem appears to be not at all trivial and requires
an application of rigorous methods of Quantum Electrodynamics (QED) for the bound
electrons.

The problem of the two-photon transitions with cascades was first discussed
in \cite{Drake}, \cite{Savukov} where the calculations were performed for the
E1M1 transitions in the He-like Highly Charged Ions (HCI). In these works the
Lorentzian fit was employed for the description of the cascade contribution to
the total two-photon frequency distribution. A rigorous QED approach for the
evaluation of the two-photon decay probability in presence of the cascades was
developed in \cite{LabShon2004}. This approach was based on QED theory of the
spectral line profile theory \cite{Low}, \cite{AndrLab} and was called Line
Profile Approach (LPA). With this method the calculations in
\cite{LabShon}-\cite{LabSol} and partly in \cite{Amaro} were performed.
An ``alternative'' approach to the cascade problem was suggested in
\cite{Jent1}-\cite{Jent3}. In these works it was claimed
that the cascade divergency in the two-photon frequency distribution,
contrary to its treatment in the LPA, can be avoided without the introduction of
the level widths in the energy denominators. The possibility of separating
out the cascade probability from the total two-photon distribution was also claimed.
The works \cite{Jent1}, \cite{Jent3} were criticized in \cite{LSP}, where
the ambiguity of the separation the cascade and ``pure'' two-photon contributions
were demonstrated on the example of the $3s\rightarrow 1s+2\gamma(E1)$ transition
probability in hydrogen. Very recently a new paper \cite{Jent4}
did arrive; the author employs finally a prescription where the widths of the intermediate
levels are introduced as in LPA. Still the separating out of the cascade is
presumed as being feasible. To our mind, the recipe given in \cite{Jent4} for
this separation is again ambiguous. Moreover, the interference between the cascade
contribution and ``pure'' two-photon contribution is absent in \cite{Jent4}. This
interference was evaluated explicitly in \cite{LSP}, \cite{LabShon} and was shown
to be comparable with the ``pure'' two-photon contribution, thus making the
attempts to separate out the cascade contribution superfluous. The reasons, why these two
contributions are comparable, are discussed in \cite{LabSolPRA}.

Recently the two-photon E1E1, E1E2, E1M1 and M1M1, E2E2 decays rates to $1s$ ground
state for the highly excited $ns$, $nd$ and $np$ states in hydrogen atom were
evaluated \cite{BeyDip}. The two goals were pursued: first, the comparison of
the nonrelativistic dipole approximation with fully relativistic evaluation.
For highly excited states the dipole approximation could become invalid: if
the arguments of the Bessel functions in the exact relativistic photon emission
operator are not small (due to the larger values for the Bohr radius of the
highly excited orbits) the dipole approximation should break down. Accordingly,
one of the conclusions in \cite{BeyDip} was the satisfactory accuracy of the
nonrelativistic dipole approximation for the highly excited states decays to
the ground state. This can be explained by the presence of the short ranged $1s$
wave function in the expression for these transition rates; the large values of
the Bessel function arguments do not affect the transition matrix elements. The
second goal in \cite{BeyDip} was to show that two-photon transition probabilities
decrease with increasing of the principal quantum number of the initial state.
In principle, this conclusion allows to restrict the number of highly excited
states decays for astrophysical tasks within the desirable accuracy.

In \cite{LabSolPRA} a theory which takes into account the ``pure'' two-photon
and cascade contributions for the two-photon and the multiphoton decays in
hydrogen was formulated. In this theory only two types of the level decays
should be present: the direct one-photon decays when they are allowed and
the total two-photon decays without separation of the ``pure'' two-photon
decays and cascades. All the decays of the excited levels should be classified
and described either as direct one-photon transitions to the ground state
or as two-photon transitions with cascades. In the ``two-photon'' approximation
transitions with more than two nonresonant photons should be neglected.
In \cite{LabSolPRA} it was demonstrated that the rigorous QED treatment of
the $3p$ level decay should include the two-photon contribution comparable with
the widely discussed two-photon decay of $3s$ level. 

In many papers on the two-photon transitions cited above it is important to
know how to correlate the transition rates between two multiplets evaluated in
the fully nonrelativistic approximation with the transition rates between the
separate components of the two multiplets, evaluated either within the Pauli
approximation (i.e. introducing the spin-orbit coupling) or with the fully
relativistic approach, where transitions between the separate components arise
automatically. For the one-photon transitions, within the $LS$-coupling scheme the
first step towards the understanding of this correlation was made in \cite{Burger,Orns},
where it was discovered empirically that for the transition between
two arbitrary multiplets $\gamma LSJ\rightarrow \gamma'L'S'J'$ within $LS$-coupling
scheme the sum of the strengths of the lines having a given initial level is
proportional to the statistical weight $(2J+1)$ of that initial level
\begin{eqnarray}
\label{intr.1}
\sum\limits_{J'}S(\gamma LSJ, \gamma'L'S'J')=(2J+1)F(\gamma L, \gamma'L')\,.
\end{eqnarray}
The line strength $S(\gamma LSJ, \gamma'L'S'J')$ is related to transition rate
$W(\gamma LSJ, \gamma'L'S'J')$ via
\begin{eqnarray}
S(\gamma LSJ, \gamma'L'S'J')=(2J+1)W(\gamma LSJ, \gamma'L'S'J').
\end{eqnarray}
The quantum numbers $L,S,J$ correspond to the total orbital momentum, total spin momentum and total angular momentum of an atom. The other quantum numbers are denoted by the symbol $\gamma$. For the allowed transitions in the nonrelativistic approximation $S'=S$ and $L'=L,L\pm 1$. The function $F$ in the right-hand side of Eq. (\ref{intr.1}) does not depend on $J$. In the book by Condon and Shortley \cite{Short} this sum rule was proven theoretically and generalized by summation of both sides of Eq. (\ref{intr.1}) over $J$. Then, remembering that $\sum\limits_{J}=(2S+1)(2L+1)$ it follows from Eq. (\ref{intr.1}):
\begin{eqnarray}
\label{intr.3}
\sum\limits_{J'J}S(\gamma LSJ, \gamma'L'S'J')=(2S+1)(2L+1)F(\gamma L, \gamma'L').
\end{eqnarray}
Later in the book by Sobelman \cite{Sobelman} it was noticed that $F(\gamma L, \gamma' L')=W(\gamma L, \gamma' L')$ where $W(\gamma L, \gamma' L')$ is the transition probability between two levels $\gamma LS$ and $\gamma' L'S'$ in the fully nonrelativistic approximation and the correlation formulated above was given in the form:
\begin{eqnarray}
\label{intr.4}
\sum\limits_{J'J}\frac{1}{2J+1}W(\gamma LSJ, \gamma'L'S'J')=(2S+1)(2L+1)W(\gamma L, \gamma'L').
\end{eqnarray}
This is the formula, which we want to generalize to the case of the multiphoton transitions. Though our proof will be given for the one-electron atom, its generalization to the many-electron atoms within the $LS$-coupling scheme seems straightforward.

\section{Multiplet sum rule for the one-photon transitions}

To start we give the proof of the formula (\ref{intr.4}) within the QED approach. The standard definition of the transition rate in the $S$-matrix formalism will be used. The $S$-matrix element for the one-photon emission process in the one-electron atom is
\begin{eqnarray}
\langle n_f j_fl_f m_{j_f}|\hat{S}^{(1)}|n_i j_i l_i m_{j_i}\rangle = e \int d^4 x\, \bar{\psi}_{n_f j_f l_f m_{j_f}}(x)\gamma_{\mu}A^*_{\mu}(x)\psi_{n_i j_i l_i m_{j_i}}(x)\,.
\end{eqnarray}
Here $\hat{S}^{(1)}$ is the first-order $S$-matrix, $e$ is the electron charge, $n$ is the principal quantum number, $j$, $m$ are the total electron angular momentum and its projection, $l$ is the orbital momentum of the corresponding state. $\psi_{n_i j_i l_i m_{j_i}}(x) = \psi_{n_i j_i l_i m_{j_i}}(\vec{r})e^{-i E_{n_i j_i l_i} t}$, $\psi_{n_i j_i l_i m_{j_i}}(\vec{r})$ is the solution of the Dirac equation for the atomic electron, $E_{n_i j_i l_i}$ is the Dirac energy, $\bar{\psi}_{n_f j_f l_f m_{j_f}} = \psi_{n_f j_f l_f m_{j_f}}^\dagger \gamma_0$ is the Dirac conjugated wave function with $\psi_{n_f j_fl_f}^{\dagger}$ being its Hermitian conjugate, 
$\gamma_{\mu} = (\gamma_0, \vec\gamma)$ are the Dirac matrices and $x\equiv (\vec{r},\,t)$ is the coordinate 4-vector ($\vec{r},\, t$ are the space- and time-coordinates). The photon field, or the photon wave function $A_{\mu}(x)$ looks like
\begin{eqnarray}
\label{2}
A^{(\vec e,\,\vec k)}_{\mu}(x) = \sqrt{\frac{2\pi}{\omega}}\,e^{(\lambda)}_{\mu}e^{i(\vec{k}\vec{r}-\omega t)}
 = \sqrt{\frac{2\pi}{\omega}}e^{-i\omega t}\,A^{(\vec e,\,\vec k)}_{\mu}(\vec r\,)
\, ,
\end{eqnarray}
where $e^{(\lambda)}_{\mu}$ is the photon polarization 4-vector, $k=(\vec{k},\omega)$ is the photon momentum 4-vector ($\vec{k}$ is the photon wave vector, $\omega=|\vec{k}|$ is the photon frequency).

After the time integration the transition amplitude $U_{n_f j_fl_f;n_i j_i l_i}^{(1\gamma)}$ is defined as
\begin{eqnarray}
\label{3}
\langle n_f j_f l_f m_{j_f}|\hat{S}^{(1)}|n_i j_i l_i m_{j_i}\rangle = -2\pi\, i\delta\left(\omega-E_{n_i j_i l_i}+E_{n_f j_fl_f}\right)U_{n_f j_f l_f m_{j_f};n_i j_i l_i m_{j_i}}^{(1\gamma)}\, .
\end{eqnarray}
Transition probability per time unit (transition rate) is defined via $U_{n_f j_f l_f m_{j_f} ;n_i j_i l_i m_{j_i}}$ like
\begin{eqnarray}
\label{4}
W_{n_f j_f l_f;n_i j_i l_i}^{(1\gamma)}= 2\pi\frac{1}{2j_i+1}\sum\limits_{m_{j_f}m_{j_i}}\left|U_{n_f j_f l_f m_{j_f};n_i j_i l_i m_{j_i}}^{(1\gamma)}\right|^2\delta\left(\omega-E_{n_i j_i l_i}+E_{n_f j_fl_f}\right)\, .
\end{eqnarray}

If the final state belongs to the continuous spectrum (as in our case due to the emitted photon) the differential transition probability should be introduced:
\begin{eqnarray}
\label{5}
dW_{n_f j_f l_f;n_i j_i l_i}^{(1\gamma)}(\vec{k},\vec{e})=2\pi\sum\limits_{m_{j_f} m_{j_i}}\frac{1}{2j_i+1}\left|U_{n_f j_f l_f m_{j_f};n_i j_i l_i m_{j_i}}^{(1\gamma)}\right|^2\delta\left(\omega-E_A+E_{A'}\right)
\frac{d\vec k}{(2\pi)^3}\, ,
\end{eqnarray}
where $d\vec k\equiv d^3k = \omega^2 d\vec{\nu}d\omega $, $d\vec{\nu}$ is the element of the solid angle in the momentum space. 
Integration in Eq. (\ref{6}) over $\omega$ gives the probability of the photon emission with polarization $\vec{e}$ in the direction $\vec{\nu}\equiv\vec{k}/\omega$ per time unit within solid angle $d\vec{\nu}$:
\begin{eqnarray}\label{6}
dW_{n_f j_fl_f;n_i j_i l_i}^{(1\gamma)}=\frac{e^2}{2\pi}\sum\limits_{m_{j_f}m_{j_i}}\frac{1}{2j_i+1}\omega_{n_f j_fl_f;n_i j_i l_i}\left|\left(( \vec{e}^{\,*}\vec{\alpha})e^{-i\vec{k}\vec{r}}\right)_{n_f j_f l_f m_{j_i};n_i j_i l_im_{j_i}}\right|^2d\vec{\nu}\, ,
\end{eqnarray}
where $\omega_{n_f j_fl_f;n_i j_i l_i}=E_{n_i j_i l_i}-E_{n_f j_fl_f}$. The total transition probability follows from Eq. (\ref{7}) after integration over angles and summation over the polarizations
\begin{eqnarray}
\label{7}
W_{n_f j_fl_f;n_i j_i l_i}^{(1\gamma)}=\frac{e^2}{2\pi}\omega_{n_f j_fl_f;n_i j_i l_i}\sum\limits_{m_{j_f}m_{j_i}}\frac{1}{2j_i+1}\sum\limits_{\vec{e}}\int d\vec{\nu}\left|\left(( \vec{e}^{\,*}\vec{\alpha})e^{-i\vec{k}\vec{r}}\right)_{n_f j_f l_f m_{j_f};n_i j_i l_i m_{j_i}}\right|^2
\end{eqnarray}

Formula (\ref{7}) describes the transitions between the fine structure components of the atomic levels. If we characterize initial and final states in fully nonrelativistic way, i.e. by the principal quantum number $n$, orbital momentum $l$ and its projection $m_l$, the transition probability is given by
\begin{eqnarray}
\label{9}
W_{n_f l_f;n_i l_i}^{(1\gamma)}=\frac{e^2}{2\pi}\frac{1}{2l_i+1}\omega_{n_f l_f;n_i l_i}\sum\limits_{m_{l_i}m_{l_f}}\sum\limits_{\vec{e}}\int d\vec{\nu}\left|\left(( \vec{e}^{\,*}\vec{\alpha})e^{-i\vec{k}\vec{r}}\right)_{n_f l_f m_{l_f};n_i l_i m_{l_i}}\right|^2\, .
\end{eqnarray}

The multiplet transition sum rule Eq. (\ref{intr.4}) in case of the one-electron atom looks like
\begin{eqnarray}
\label{10}
W_{n_f l_f;n_i l_i}^{(1\gamma)}=\frac{1}{(2l_i+1)(2s+1)}\sum\limits_{j_ij_f}(2j_i+1)W_{n_f j_fl_f;n_i j_i l_i}^{(1\gamma)}\, ,
\end{eqnarray}
where $s$ is the electron spin ($s=1/2$).

This formula in the Pauli approximation can be easily proved after presentation of the wave function $\psi_{njm}$ in the form
\begin{eqnarray}
\label{11}
\psi_{n j l m}(\vec{r})=\sum\limits_{m_l m_s}C^{jm}_{lm_l\,\,sm_s}R_{nl}(r)Y_{lm_l}\chi_{sm_s}\, .
\end{eqnarray}
Here $C^{jm}_{lm_l\,\,sm_s}$ are the Clebsch-Gordan coefficients (see,for example, \cite{Varsh}), $R_{nl}(r)$ is the radial part of the wave function, $Y_{lm_l}$ is the spherical function describing angular part of the wave function and $\chi_{sm_s}$ is the spin part of the nonrelativistic Schr\"odinger wave function.

Using Eq. (\ref{10}), the expression for the transition probability Eq. (\ref{9}) can be rewritten like
\begin{eqnarray}
\label{12}
W_{n_f j_fl_f;n_i j_i l_i}^{(1\gamma)}=\frac{e^2\omega_{n_f j_fl_f;n_i j_i l_i}}{2\pi(2j_i+1)}\sum\limits_{m_im_f}\sum\limits_{\vec{e}}\int d\vec{\nu}\left[\sum\limits_{m_{l_i}m_{l_f}}\sum\limits_{m_{s_i}m_{s_f}}C^{j_im_i}_{l_im_{l_i}\,\,s_im_{s_i}}C^{j_fm_f}_{l_fm_{l_f}\,\,s_fm_{s_f}}\delta_{s_is_f}\delta_{m_{s_i}m_{s_f}}\sum\limits_{m'_{l_i}m'_{l_f}}\times
\right.
\\
\nonumber
\left.
\sum\limits_{m'_{s_i}m'_{s_f}}C^{j_im_i}_{l_im'_{l_i}\,\,s'_im'_{s_i}}C^{j_fm_f}_{l_fm'_{l_f}\,\,s_fm'_{s_f}}\delta_{s_is_f}\delta_{m'_{s_i}m'_{s_f}}
\langle n_il_im_{l_i}|( \vec{e}^{\,*}\vec{\alpha})e^{-i\vec{k}\vec{r}}|n_fl_fm_{l_f}\rangle \langle n_il_im'_{l_i}|( \vec{e}^{\,*}\vec{\alpha})e^{-i\vec{k}\vec{r}}|n_fl_fm'_{l_f}\rangle^*\right],
\end{eqnarray}
where the wave function $\langle n_il_im_{l_i}|\equiv R_{n_il_i}(r)Y^*_{l_im_i}$. In Eq. (\ref{12}) the ortho-normalization of the spin functions is used and $\delta_{ik}$ is the Kronecker symbol. Now we can perform the summation over $j_ij_f$ and $m_im_f$ for the Clebsch-Gordan coefficients (see \cite{Varsh}):
\begin{eqnarray}
\label{13}
\sum\limits_{j_im_i}\sum\limits_{j_fm_f}C^{j_im_i}_{l_im_{l_i}\,\,s_im_{s_i}}C^{j_fm_f}_{l_fm_{l_f}\,\,s_fm_{s_f}}C^{j_im_i}_{l_im'_{l_i}\,\,s_im'_{s_i}}C^{j_fm_f}_{l_fm'_{l_f}\,\,s_fm'_{s_f}}=\delta_{m_{l_i}m'_{l_i}}\delta_{m_{s_i}m'_{s_i}}\delta_{m_{l_f}m'_{l_f}}\delta_{m_{s_f}m'_{s_f}}\, .
\end{eqnarray}
Now summation of Eq. (\ref{12}) over $j_f$, $j_i$ with coefficient $(2j_i+1)$ and the use of Eq. (\ref{9}) result
\begin{eqnarray}
\label{14}
\sum\limits_{j_ij_f}(2j_i+1)W_{n_f j_fl_f;n_i j_i l_i}^{(1\gamma)}=\frac{e^2}{2\pi}\sum\limits_{m_{l_i}m_{l_f}}\sum\limits_{m_{s_i}m_{s_f}}\omega_{n_fl_f;n_il_i}\sum\limits_{\vec{e}}\int d\vec{\nu}\left|\left(( \vec{e}^{\,*}\vec{\alpha})e^{-i\vec{k}\vec{r}}\right)_{n_f m_{l_f};n_i m_{l_i}}\right|^2\delta_{m_{s_i}m_{s_f}}\, .
\end{eqnarray}
Using the equality $\sum\limits_{m_{s_i}m_{s_f}}\delta_{m_{s_i}m_{s_f}}=(2s_i+1)$, we obtain finally
\begin{eqnarray}
\label{15}
\sum\limits_{j_ij_f}(2j_i+1)W_{n_f j_fl_f;n_i j_i l_i}^{(1\gamma)}=(2l_i+1)(2s_i+1)W_{n_f l_f;n_i l_i}^{(1\gamma)}\, ,
\end{eqnarray}
which proves the sum rule Eq. (\ref{10}).

\section{Multiplet sum rule for the multiphoton transitions}

In this section we prove the multiplet sum rule for the multiphoton transitions on the example of the two-photon emission process. We use again the Pauli approximation for the wave functions, but simplify the expressions for the transition operators. Here we present these operators in the fully nonrelativistic limit what does not influence the proof.

The transition probability for the two-photon emission process can be written in the form \cite{AndrLab}:
\begin{eqnarray}
\label{16}
dW_{n_fj_fl_f;n_ij_il_i}^{(2\gamma)}(\omega',\vec{\nu},\vec{\nu}\,',
\vec{e},\vec{e}\,')=e^4
\frac{\omega'(E_{n_il_i}-E_{n_fl_f}-\omega')}{(2\pi)^3}\sum\limits_{m_{i}m_{f}}\frac{1}{2j_{i}+1}\times
\nonumber\\
\left|\sum_{njlm}\frac{\langle n_fj_fl_fm_f|\vec{e}^{\,*}\vec{r}|njlm\rangle \langle njlm |\vec{e}\,'^{\,*}\vec{r}|n_ij_il_im_i\rangle}{E_{nl}-E_{n_il_i}+\omega'}+\sum_{njlm}\frac{\langle n_fj_fl_fm_f|\vec{e}\,'^{\,*}\vec{r}|njlm\rangle \langle njlm |\vec{e}^{\,*}
\vec{r}|n_ij_il_im_i\rangle}{E_{nl}-E_{n_il_i}+\omega}\right|^2d{\vec{\nu}}
d\vec{\nu}\,'d\omega'\, ,
\end{eqnarray}
where summation extends over the total set of the wave functions for the intermediate states in the Pauli approximation. In Eq. (\ref{16}) we neglect the fine structure, i.e. energy levels do not depend on $j$. Consider only the first term in Eq. (\ref{16}), using again function Eq. (\ref{11}):
\begin{eqnarray}\label{3.2}
\sum\limits_{njlm}\frac{\langle n_fj_fl_fm_f|\vec{e}^{\,*}\vec{r}|njlm\rangle \langle njlm |\vec{e}\,'^{\,*}\vec{r}|n_ij_il_im_i\rangle}{E_n-E_{n_il_i}+\omega'}=\sum_n\sum\limits_{jm lm_{l}}\sum\limits_{m_{l_i}m_{l_f}}\sum\limits_{m_{s_i}m_{s_f}}\sum\limits_{m_{l}m'_{l}}\sum\limits_{m_{s}m'_{s}}\times
\nonumber
\\
\delta_{s_is}\delta_{ss_f}\delta_{m_{s_i}m'_s}\delta_{m_sm_{s_f}}C^{j_im_i}_{l_im_{l_i}\,\,s_im_{s_i}}C^{j_fm_f}_{l_fm_{l_f}\,\,s_fm_{s_f}}C^{jm}_{lm_{l}\,\,sm_{s}}C^{jm}_{lm'_{l}\,\,sm'_{s}}
\frac{\langle n_fl_fm_{l_f}|\vec{e}^{\,*}\vec{r}|nlm_{l}\rangle 
\langle nlm'_{l}|\vec{e}\,'^{\,*}\vec{r}|n_il_im_{l_i}\rangle }{E_n-E_{n_il_i}+\omega'}.
\end{eqnarray}
The second term in Eq. (\ref{16}) can be calculated in the same way. The summation over quantum numbers $jm$ in Eq. (\ref{3.2}) can be performed by $\sum\limits_{jm}C^{jm}_{lm_{l}\,\,sm_{s}}C^{jm}_{lm'_{l}\,\,sm'_{s}}=\delta_{m_lm'_l}\delta_{m_sm'_s}$. Squared expression (\ref{3.2}) can be written in the form:
\begin{eqnarray}
\label{3.3}
\left|\sum\limits_{njlm}\frac{\langle n_fj_fl_fm_f|\vec{e}^{\,*}\vec{r}|njlm\rangle \langle njlm |\vec{e}\,'^{\,*}\vec{r}|n_ij_il_im_i\rangle}{E_n-E_{n_il_i}+\omega'}\right|^2=\sum\limits_{nn'}\sum\limits_{m_{l_i}m'_{l_i}}\sum\limits_{m_{l_f}m'_{l_f}}\sum\limits_{m_sm'_s}\sum\limits_{m_lm'_l}C^{j_im_i}_{l_im_{l_i}\,\,sm_s}C^{j_fm_f}_{l_fm_{l_f}\,\,sm_s}
\nonumber
\\
C^{j_im_i}_{l_im'_{l_i}\,\, sm'_s}C^{j_fm_f}_{l_fm'_{l_f}\,\,sm'_s}\frac{
\langle n_fl_fm_{l_f}|\vec{e}^{\,*}\vec{r}|nlm_{l}\rangle 
\langle n_fl_fm'_{l_f}|\vec{e}^{\,*}\vec{r}|nlm'_{l}\rangle ^*
\langle nlm_{l}|\vec{e}\,'^{\,*}\vec{r}|n_il_im_{l_i}\rangle
\langle nlm'_{l}|\vec{e}\,'^{\,*}\vec{r}|n_il_im_{l_i}\rangle ^*}
{(E_n-E_{n_il_i}+\omega')(E_{n'}-E_{n_il_i}+\omega')}.
\end{eqnarray}
Then for the first term of the two-photon transition probability we can write
\begin{eqnarray}
\label{3.4}
\sum\limits_{j_ij_f}(2j_i+1)dW_{n_fj_fl_f;n_ij_il_i}^{(2\gamma)a}(\omega',\vec{\nu},\vec{\nu}\,',
\vec{e},\vec{e}\,')=e^4
\frac{\omega'(E_{n_il_i}-E_{n_fl_f}-\omega')}{(2\pi)^3}\times
\nonumber
\\
\sum\limits_{j_ij_f}\sum\limits_{m_{i}m_{f}}\sum\limits_{nn'}\sum\limits_{m_{l_i}m'_{l_i}} \sum\limits_{m_{l_f}m'_{l_f}}\sum\limits_{m_sm'_s}\sum\limits_{m_lm'_l}C^{j_im_i}_{l_im_{l_i}\,\,sm_s}C^{j_fm_f}_{l_fm_{l_f}\,\,sm_s}C^{j_im_i}_{l_im'_{l_i}\,\, sm'_s}C^{j_fm_f}_{l_fm'_{l_f}\,\,sm'_s}\times
\\
\nonumber
\frac{
\langle n_fl_fm_{l_f}|\vec{e}^{\,*}\vec{r}|nlm_{l}\rangle 
\langle n_fl_fm'_{l_f}|\vec{e}^{\,*}\vec{r}|nlm'_{l}\rangle ^*
\langle nlm_{l}|\vec{e}\,'^{\,*}\vec{r}|n_il_im_{l_i}\rangle
\langle nlm'_{l}|\vec{e}\,'^{\,*}\vec{r}|n_il_im_{l_i}\rangle ^*}
{(E_n-E_{n_il_i}+\omega')(E_{n'}-E_{n_il_i}+\omega')}d{\vec{\nu}}
d\vec{\nu}\,'d\omega'
\end{eqnarray}
Now we can perform summation over $j_im_i$ and $j_fm_f$, which, together with the second term in (\ref{16}), leads to
\begin{eqnarray}
\label{3.5}
\sum\limits_{j_ij_f}(2j_i+1)dW_{n_fj_fl_f;n_ij_il_i}^{(2\gamma)}(\omega',\vec{\nu},\vec{\nu}\,',
\vec{e},\vec{e}\,')=e^4
\frac{\omega'(E_{n_il_i}-E_{n_fl_f}-\omega')}{(2\pi)^3}\times
\nonumber
\\
(2s_i+1)\sum\limits_{m_{l_i}m_{l_f}}\left|\sum\limits_{nlm_l}\frac{
\langle n_fl_fm_{l_f}|\vec{e}^{\,*}\vec{r}|nlm_{l}\rangle 
\langle nlm_{l}|\vec{e}\,'^{\,*}\vec{r}|n_il_im_{l_i}\rangle}
{E_n-E_{n_il_i}+\omega'} +
\sum\limits_{nlm_l}\frac{
\langle n_fl_fm_{l_f}|\vec{e}\,'^{\,*}\vec{r}|nlm_{l}\rangle 
\langle nlm_{l}|\vec{e}^{\,*}\vec{r}|n_il_im_{l_i}\rangle}
{E_n-E_{n_il_i}+\omega}\right|^2.
\end{eqnarray}
Multiplying now Eq. (\ref{3.5}) by the factor $\frac{2l_i+1}{2l_i+1}$, summing over polarizations $\vec{e}\,,\vec{e}\,'$, integrating over photon directions and over $\omega'$, we get
\begin{eqnarray}
\label{3.6}
\sum\limits_{j_ij_f}(2j_i+1)W_{n_fj_fl_f;n_ij_il_i}^{(2\gamma)}=(2s_i+1)(2l_i+1)W_{n_fl_f;n_il_i}^{(2\gamma)}.
\end{eqnarray}
This equality proves the sum rule for the two-photon transition rates between multiplets. In principle, such evaluation can be easily extended to the case of the multiphoton transitions. The final expression can be presented like
\begin{eqnarray}
\label{3.7}
W_{n_fl_f;n_il_i}^{(k\gamma)}=\frac{1}{(2s_i+1)(2l_i+1)}\sum\limits_{j_ij_f}(2j_i+1)W_{n_fj_fl_f;n_ij_il_i}^{(k\gamma)}\, ,
\end{eqnarray}
where $k$ denotes the number of the emitted photons.

\section{Testing the multiplet sum rule for the two-photon transitions}

We will test the multiplet transition sum rule on the E1E1 two-photon emission processes. The standard expression for the transition probability for the two-photon emission processes is (see e.g. \cite{AndrLab})
\begin{eqnarray}
\label{19}
dW_{AA'}^{(2\gamma)}(\omega',\vec{\nu},\vec{\nu}\,',
\vec{e},\vec{e}\,')=e^4
\frac{\omega'(E_A-E_{A'}-\omega')}{(2\pi)^3}\frac{1}{2j_{A}+1}\sum\limits_{m_{A}m_{A'}}
\nonumber\\
\times\left|\sum_N\frac{({\vec{\alpha}}\vec{A}^*_{\vec{e},
\vec{k}})_{A'N} ({\vec{\alpha}}\vec{A}^*_{\vec{e}\,',
\vec{k}'})_{NA}}{E_N-E_A+\omega'}+\sum_N\frac{({\vec{\alpha}}\vec{A}^*_{\vec{e}\,',
\vec{k}'})_{A'N} ({\vec{\alpha}}\vec{A}^*_{\vec{e},
\vec{k}})_{NA}}{E_N-E_A+\omega}\right|^2d{\vec{\nu}}
d\vec{\nu}\,'d\omega'\, ,
\end{eqnarray}
where, as usual, $A$ $A'$ $N$ denote the set of quantum numbers $n j l m$ or $n l m_l$, for the relativistic or nonrelativistic cases, respectively. Notations $A$ $A'$ $N$ correspond to the initial, final or intermediate states.

Using again the nonrelativistic expressions for the transition operators and after the summation over polarizations and integration over photons directions for the E1E1 two-photon emission we obtain
\begin{eqnarray}\label{20}
dW^{\rm E1E1}_{A\rightarrow
A'}(\omega_2)&=&\frac{8}{9\pi}\left(\frac{4\pi}{3}\right)^2\frac{1}{2j_A+1}\sum\limits_{m_im_f}\sum\limits_{qq'=0,\pm 1}(-1)^{q+q'}\left|\langle A'|rY_{1q}\left({\bf
n}_{\bf r}\right)G(E_{A}-\omega;{\bf r},{\bf
r}')r'Y_{1q'}^*\left({\bf n}_{\bf r'}\right)|A\rangle\right.\nonumber
\\
&& + \left.\langle A'|rY_{1q'}\left({\bf n}_{\bf r}\right)G(E_{A}-\omega';{\bf r}, {\bf r}\,')r'Y^*_{1q}\left({\bf n}_{\bf r'}\right)|A\rangle\right|^2 (\omega\omega')^3\,d\omega\, ,
\end{eqnarray}
where  $\vec{r}_q=\sqrt{\frac{4\pi}{3}}rY_{1q}$, $\omega'\equiv E_A-E_{A'}-\omega$ and $G(E;{\bf r}, {\bf r}\,')$ is the Coulomb Green function. With the use of the Green function partial wave decomposition
\begin{eqnarray}\label{21}
G(E;{\bf r},{\bf
r}')=\sum\limits_{lm_l}\frac{1}{rr'}\,g_l(E;r,r')\,Y_{lm_l}^*({\bf
n}_{\bf r})Y_{lm_l}({\bf n}_{\bf r'})\, ,
\end{eqnarray}
it is easy to perform angular integration for the required process.

The simplest situation occurs for the $3d_{j_i}\rightarrow 2\gamma(E1)+2s_{j_f}$ transition in the nonrelativistic limit. In this case the intermediate states are $np_{j_n}$ states only. After the angular integration and summation over all projections in Eq. (\ref{20}) the differential transition probability can be presented like
\begin{eqnarray}
\label{24}
dW_{3d_{5/2}2s_{1/2}}^{\rm E1E1}=\frac{8\omega^3\omega'^3}{54\pi}\left(\frac{2}{\sqrt{5}}I_1(E_{A}-\omega)+\frac{2}{\sqrt{5}}I_1(E_{A}-\omega')\right)^2
\end{eqnarray}
and
\begin{eqnarray}
\label{25}
dW_{3d_{3/2}2s_{1/2}}^{\rm E1E1}=\frac{8\omega^3\omega'^3}{36\pi}\left(2\sqrt{\frac{2}{15}}I_1(E_{A}-\omega)+2\sqrt{\frac{2}{15}}I_1(E_{A}-\omega')\right)^2\, ,
\end{eqnarray}
where 
\begin{eqnarray}
I_1(E_{A}-\omega)\equiv \int\limits_0^{\infty}\int\limits_0^{\infty}dr_1dr_2r^3r^3R_{20}(r_1)g_1(E_{A}-\omega;r_1,r_2)R_{32}(r_2)\,.
\end{eqnarray}
This result corresponds to the Pauli approximation with the neglect of the energy splitting of the fine structure components. Nonrelativistic evaluation without consideration of the separate fine structure components gives
\begin{eqnarray}
\label{26}
dW_{3d\,2s}^{\rm E1E1}=\frac{8\omega^3\omega'^3}{45 \pi}\left(\sqrt{\frac{2}{3}}I_1(E_{A}-\omega)+\sqrt{\frac{2}{3}}I_1(E_{A}-\omega')\right)^2\, .
\end{eqnarray}
The integration over radial variables can be easily done with the radial Coulomb Green function decomposition over Laguerre polynomials \cite{RZM}:
\begin{eqnarray}\label{3.55}
g_l(\nu; r,r')=\frac{4Z}{\nu}\left(\frac{4}{\nu^2}rr'\right)^l
\exp\left(-\frac{r+r'}{\nu}\right)\sum\limits_{n=0}^{\infty}\frac{n!L^{2l+1}_n\left(\frac{2r}{\nu}\right)L^{2l+1}_n\left(\frac{2r'}{\nu}\right)}{(2l+1+n)!(n+l+1-\nu)}.
\end{eqnarray}
Thus the radial integrals are the same for the calculations with or without fine structure. 

The relativistic expression for the transition probability is
\begin{eqnarray}
\label{22}
dW_{AA'}^{\rm E1E1}(\omega) = e^4
\frac{32\pi\,\omega\,\omega'}{2j_{A}+1}
\sum\limits_{M\,M'\,m_{A}\,m_{A'}}
\Biggl|\sum_N\frac{(A^{(1)}_{1M})_{A'N}
(A^{(1)}_{1M'})_{NA}}{E_N-E_A+\omega'}
+\sum_N\frac{(A^{(1)}_{1M'})_{A'N}
(A^{(1)}_{1M})_{NA}}{E_N-E_A+\omega}\Biggr|^2 d\omega\, ,
\end{eqnarray}
with
\begin{eqnarray}
\label{phot_len}
A^{(1)}_{JM}(\omega)&=&
-j_{J+1}(\omega r)\ \vec{\alpha}\vec{Y}^{(1)}_{JM}(\hr)
+\sqrt\frac{J+1}{J}j_{J+1}(\omega r)\ \vec{\alpha}\vec{Y}^{(-1)}_{JM}(\hr)
\nonumber \\
&&-i\sqrt\frac{J+1}{J}j_J(\omega r)Y_{JM}(\hr){\rm I}\,,
\end{eqnarray}
where $Y_{JM}(\hr)$ are the spherical functions and $j_J(\omega r)$ is the Bessel's function.
Explicit formulas for the one-electron matrix elements $A^{(\lambda)}_{JM}(\omega)$ in the length and velocity gauges can be found in \cite{John}, \cite{LabShonSol}. We calculate all transition rates in nonrelativistic limit by the Coulomb Green function method and in relativistic way by the dual-kinetic-balance finite basis set method \cite{shabaev:2004:130405}.

In Tables~\ref{tab:1},~\ref{tab:2},~\ref{tab:3},~\ref{tab:4} the numerical values of the two-photon E1E1 transition probabilities for the $3p_{j_i}\rightarrow 2\gamma(E1)+2p_{j_f}$, $4p_{j_i}\rightarrow 2\gamma(E1)+3p_{j_f}$, $3d_{j_i}\rightarrow 2\gamma(E1)+2s_{j_f}$ and $4d_{j_i} \rightarrow 2\gamma(E1)+ 3d_{j_f}$ are presented.

The values of the nonrelativistic and relativistic calculations in Table~\ref{tab:1} are in perfect agreement with each other. The small difference is due to the energies in the denominators of the transition amplitude Eq. (\ref{19}). As it was mentioned before in the nonrelativistic calculations the energies of all the fine structure components are equal. We would note also that the relativistic calculations are in perfect agreement with the ones in \cite{Amaro}. Apart from the $3d-2s$ transition we have evaluated also $W^{(E1E1)}_{4p_{j_i}-3p_{j_f}}$, $W^{(E1E1)}_{4d_{j_i}-3d_{j_f}}$ and $W^{(E1E1)}_{3d_{j_i}-2s_{j_f}}$ two-photon transition probabilities.

\section{Conclusions}
In our paper we have considered the processes of two-photon transitions for hydrogenic atom.
Recent astrophysical investigations necessitate a detailed analysis of the multiphoton
emission processes and, most important, of the two-photon radiation. The ``pure'' two-photon
emission leads to the photon escape from matter and, thus, presents the formation
mechanism for the background radiation. In the recent paper \cite{LabSolPRA} we have
developed the ``two-photon'' approximation method for the intricate multi-photon emission
processes. As the extension of this paper we considered two-photon E1E1 transitions
between neighboring $3p-2p$, $4p-3p$, $3d-2s$ and $4d-3d$ atomic levels. 

We were comparing the nonrelativistic values with the relativistic ones. However,
this comparison was complicated by the the fine structure splitting. The multiplet
transition sum rule, known for the one-photon transitions, required the proof for
the multi-photon processes.

The values of the relativistic and nonrelativistic calculations with and without fine
structure splitting are presented in
Tables~\ref{tab:1},~\ref{tab:2},~\ref{tab:3},~\ref{tab:4} for the various processes.
The corresponding nonrelativistic and relativistic  values of the two-photon transition
probabilities are in perfect agreement. The slight difference occurs due to the
nonrelativistic Schr\"odinger approximation which does not take into account the
fine structure splitting.

The most important conclusion is that the multiplet-transition sum rule is correct also
for the two-photon (multiphoton) transitions and can be used for additional
checking of the results of the numerical calculations.

\begin{center}
Acknowledgments
\end{center}
The authors acknowledge financial support from RFBR (grant Nr. 08-02-00026). The work of D.S. was supported by
the Non-profit Foundation ``Dynasty'' (Moscow). The authors acknowledge also 
the support by the Program of development of scientific potential of High School,
Ministry of Education and Science of Russian Federation (grant Nr. 2.1.1/1136 and goskontrakt $\Pi$1334).
A.V. and G.P. acknowledge financial support by DFG (grant Nr. VO 1707/1-1) and GSI.

\newpage
\begin{table}
\caption{
Transition probabilities $W^{\rm E1E1}_{3p_{j_i},\,2p_{j_f}}$ in units s$^{-1}$
for the $3p_{j_i} \rightarrow 2\gamma(E1)+ 2p_{j_f}$ emission. In the first column the angular momenta for the initial and final states are given, the second column represents the numerical values of the nonrelativistic calculations in the Pauli approximation for the transition probabilities between fine structure components. In the third column the values of the relativistic calculations are given. The fourth column represents the results of the analogous relativistic calculations \cite{Amaro}. The last but one line corresponds to the averaged according to the right hand side of the Eq. (\ref{3.7}) value.  The last line corresponds to the nonrelativistic evaluation of the two-photon E1E1 decay without taking into account the fine structure splitting (left hand side of Eq. (\ref{3.7})). All results are given for the hydrogen atom.}
\label{tab:1}
\begin{tabular}{| l | c | c| r |}
\hline \hline
$j_i-j_f$ & nonrel. $W_{3p_{j_i}2p_{j_f}}^{\rm E1E1}$ $s^{-1}$ & rel. $W_{3p_{j_i}2p_{j_f}}^{\rm E1E1}$ $s^{-1}$ &  rel. $W_{3p_{j_i}2p_{j_f}}^{\rm E1E1}$ $s^{-1}$ \cite{Amaro}  \\
 \hline\hline
$\frac{1}{2}-\frac{1}{2}$ & $0.0466033$ & $0.0466015$ & $0.0466015$ \\
$\frac{1}{2}-\frac{3}{2}$ & $0.000882617$ & $0.0008832673$ & $0.0008832671$ \\
$\frac{3}{2}-\frac{3}{2}$ & $0.0470446$ & $0.04704895$ & $0.04704893$ \\
$\frac{3}{2}-\frac{1}{2}$ & $0.000441308$ & $0.0004414501$ & $0.0004414514$ \\ \hline
averaged & $0.04748591$    & $0.0474885$ & $0.0474885$ \\ \hline
$W_{3p2p}^{\rm E1E1}$ &  $0.0474859$ & \qquad & \\
\hline \hline
\end{tabular}
\end{table}

\begin{table}
\caption{
Transition probabilities $W^{\rm E1E1}_{4p_{j_i},\,3p_{j_f}}$ in units s$^{-1}$
for the $4p_{j_i} \rightarrow 2\gamma(E1)+ 3p_{j_f}$ emission. In the first column the angular momenta for the initial and final states are given, the second column represents the numerical values of the nonrelativistic calculations in the Pauli approximation for the transition probabilities between fine structure components. In the third
column the values of the relativistic calculations are given. The last but one line corresponds to the averaged according to the right hand side of the Eq. (\ref{3.7}) value. The last line corresponds to the nonrelativistic evaluation of the two-photon E1E1 decay without taking into account the fine structure splitting (left hand side of Eq. (\ref{3.7})). All results are given for the hydrogen atom.}
\label{tab:2}
\begin{tabular}{| l | c | r |}
\hline \hline
$j_i-j_f$ & nonrel. $W_{4p_{j_i}3p_{j_f}}^{\rm E1E1}$ $s^{-1}$ & rel. $W_{4p_{j_i}3p_{j_f}}^{\rm E1E1}$ $s^{-1}$   \\
 \hline\hline
$\frac{1}{2}-\frac{1}{2}$ & $0.00253805$ & $0.00253771$   \\
$\frac{1}{2}-\frac{3}{2}$ & $3.55751\cdot 10^{-5}$& $3.56159\cdot 10^{-5}$   \\
$\frac{3}{2}-\frac{3}{2}$ & $0.00255583$ & $0.0025561093$   \\
$\frac{3}{2}-\frac{1}{2}$ & $1.77876\cdot 10^{-5}$ & $1.77981\cdot 10^{-5}$   \\ \hline
averaged & $0.00257362$    & $0.002573714$   \\ \hline
$W_{4p3p}^{\rm E1E1}$ &  $0.00257362$ & \qquad  \\
\hline \hline
\end{tabular}
\end{table}

\begin{table}
\caption{
Transition probabilities $W^{\rm E1E1}_{3d_{j_i},\,2s_{j_f}}$ in units s$^{-1}$
for the $3d_{j_i} \rightarrow 2\gamma(E1)+ 2s_{j_f}$ emission. In the first column the angular momenta of the initial and final states are given, the second column represents the numerical values of the nonrelativistic calculation in Pauli approximation for the transition probabilities between fine structure components. In the third
column the values of the relativistic calculations are given. The fourth column represents the results of the analogous relativistic calculations \cite{Amaro}. The last but one line corresponds to the averaged according to the right hand side of the Eq. (\ref{3.7}) value. The last line corresponds to the nonrelativistic evaluation of the two-photon E1E1 decay without taking into account the fine structure splitting (left hand side of Eq. (\ref{3.7})). All results are given for the hydrogen atom.}
\label{tab:3}
\begin{tabular}{| l | c | c | r |}
\hline \hline
$j_i-j_f$ & nonrel. $W_{3d_{j_i}2s_{j_f}}^{\rm E1E1}$ $s^{-1}$ & rel. $W_{3d_{j_i}2s_{j_f}}^{\rm E1E1}$ $s^{-1}$&  rel. $W_{3d_{j_i}2s_{j_f}}^{\rm E1E1}$ $s^{-1}$ \cite{Amaro} \\
 \hline\hline
$\frac{5}{2}-\frac{1}{2}$ & $0.000775914$ & $0.0007750009$ & $0.0007750004$ \\
$\frac{3}{2}-\frac{1}{2}$ & $0.000775915$ & $0.0007762451$ & $0.0007762447$ \\ \hline
averaged & $0.000775915$ & $0.000775499$  & $0.000775498$\\ \hline
$W_{3d2s}^{\rm E1E1}$ &  $0.000775914$ & \qquad &  \\
\hline \hline
\end{tabular}
\end{table}

\begin{table}
\caption{
Transition probabilities $W^{\rm E1E1}_{4d_{j_i},\,3d_{j_f}}$ in units s$^{-1}$
for the $4d_{j_i} \rightarrow 2\gamma(E1)+ 3d_{j_f}$ emission. In the first column the angular momenta of the initial and final states are given, the second column represents the numerical values of the nonrelativistic calculation in the Pauli approximation for the transition probabilities between fine structure components. In the third
column the values of the relativistic calculations are given. The last but one line corresponds to the averaged according to the right hand side of the Eq. (\ref{3.7}) value. And, finally, last line corresponds to the nonrelativistic evaluation of the two-photon E1E1 decay without taking into account the fine structure splitting (left hand side of Eq. (\ref{3.7})). All results are given for the hydrogen atom.}
\label{tab:4}
\begin{tabular}{| l | c | c | r |}
\hline \hline
$j_i-j_f$ & nonrel. $W_{4d_{j_i}3d_{j_f}}^{\rm E1E1}$ $s^{-1}$ & rel. $W_{4d_{j_i}3d_{j_f}}^{\rm E1E1}$ $s^{-1}$ \\
 \hline\hline
$\frac{3}{2}-\frac{3}{2}$ & $0.00167601$ & $0.001674919$  \\
$\frac{3}{2}-\frac{5}{2}$ & $1.08221\cdot 10^{-5}$ & $1.074580\cdot 10^{-5}$  \\
$\frac{5}{2}-\frac{5}{2}$ & $0.00167961$ & $0.001678357$  \\
$\frac{5}{2}-\frac{3}{2}$ & $0.721476\cdot 10^{-5}$ & $0.7163263\cdot 10^{-5}$  \\
 \hline
averaged & $0.001686828$ & $0.00168171$  \\ \hline
$W_{4d3d}^{\rm E1E1}$ &  $0.001686826$ & \qquad  \\
\hline \hline
\end{tabular}
\end{table}
\end{document}